\documentclass[11pt]{article}       
\usepackage{epsfig,amsmath,amssymb,multirow}     

\begin{document} 
\begin{titlepage}

\title{The minimal molecular surface}

\author{P. W. Bates$^1$, 
G. W. Wei$^{1,2}$\footnote{To whom correspondence should be addressed. E-mail:wei@math.msu.edu},
and 
Shan Zhao$^1$ \\
%\address{
$^1$Department of Mathematics \\  
Michigan State University, MI 48824, USA\\
$^2$Department of Electrical and Computer Engineering \\ 
Michigan State University, MI 48824, USA
} 
\date{\today} 
\maketitle
\begin{abstract} 
We  introduce a novel concept, the minimal molecular surface (MMS), 
as a new paradigm for the theoretical modeling of biomolecule-solvent interfaces. 
When a less polar macromolecule is immersed in a polar environment, the surface
free energy minimization occurs naturally to stabilizes the system, and leads to an 
MMS separating the macromolecule from the solvent. For a given set of atomic constraints
(as obstacles), the MMS is defined as one whose mean curvature 
vanishes away from the obstacles. An iterative procedure is proposed to compute the MMS. 
Extensive examples are given to validate the proposed algorithm and illustrate the new concept. 
We show that the MMS provides an indication to DNA-binding specificity.
The proposed algorithm represents a major step forward in minimal surface generation.

\end{abstract} 

\end{titlepage}

The stability and solubility of macromolecules, such as proteins, DNAs and RNAs, are determined 
by how their surfaces interact with solvent and/or other surrounding molecules. Therefore, the 
structure and function of macromolecules depend on the features of their molecule-solvent interfaces 
\cite{Kuhn}. Molecular surface was proposed \cite{Richards,Connolly} to describe the interfaces and has 
been applied to protein folding \cite{Spolar},  protein-protein interfaces \cite{Crowley}, protein surface 
topography \cite{Kuhn}, oral drug absorption classification \cite{Bergstrom}, DNA binding and bending 
\cite{Dragan}, macromolecular docking \cite{Jackson}, enzyme catalysis \cite{LiCata}, calculation of 
solvation energies \cite{Raschke}, and molecular dynamics \cite{Das}. It is of paramount importance to 
the implicit solvent models \cite{WarWat,Honig95}. However, the molecular surface model suffers from 
it being probe dependent, non-differentiable, and being inconsistent with free energy minimization. 

Minimal surfaces are omnipresent in nature. Their study  has been a fascinating topic for centuries 
\cite{Andersson,Anderson,Pociecha}. French geometer, Meusnier, constructed the first non-trivial example, 
the catenoid, a minimal surface that connects two parallel circles, in the 18th century. In 1760, Lagrange 
discoved the relation between  minimal surfaces and a variational principle, which is still a cornerstone of 
modern mechanics. Plateau studied minimal surfaces in soap films in the mid-nineteenth century. In liquid 
phase, materials of largely different polarizabilities, such as water and oil, do not mix, and the material 
in smaller quantity forms ellipsoidal drops, whose surfaces are minimal subject to the gravitational constraint. 
The self-assembly of minimal cell membrane surfaces in water has been discussed \cite{Seddon}. The Schwarz P 
minimal surface is known to play a role in periodic crystal structures \cite{ChenEHOY}. The formation of 
$\beta$-sheet structures in proteins is regarded as the result of surface minimization on a catenoid \cite{Koh}. 
A minimal surface metric has been proposed for the structural comparison of proteins \cite{Falicov}. However,
to the best of our knowledge, a natural minimal surface that separates a less polar macromolecule from its 
polar environment such as the water solvent has not been considered yet. The objective of this Report is to 
introduce the theory of and algorithm to generate minimal molecular surfaces (MMSs). Since the surface free 
energy is proportional to the surface area, a MMS contributes to the molecular stability in solvent. Therefore, 
there must be a MMS associated with each stable macromolecule in its polar environment. Although minimal surfaces 
are often generated by evolving surfaces with predetermined curve boundaries \cite{Chopp,Cecil}, there is no 
algorithm available that generates minimal surfaces with respect to obstacles, such as atoms.  Here, we 
develop such an algorithm based on the theory of differential geometry \cite{Gray}. 
  
For a given initial function $S(x,y,z)$ that characterizes domain encompassing the biomolecule of interest,
we consider an evolution driving by the mean curvature $H$
\begin{equation}\label{evolut}
S_\varepsilon(x,y,z)=S(x,y,z)+ \varepsilon\sqrt{g}H,
\end{equation}
where $\varepsilon>0$ is a small parameter, $g=1+S_x^2+S_y^2+S_x^2$ is the Gram determinant,
and $H=\frac{1}{3} \nabla \cdot \left(\frac{\nabla S} {\sqrt{g}}\right)$.
Our procedure involves iterating Eq. (\ref{evolut}) until ${H}\sim 0$ everywhere except for 
certain protected boundary points where the mean curvatures take constant values. 
Physically, the vanishing of the mean curvature is a natural consequence of surface free
energy minimization. Consider the surface free energy of a molecule as $E=\int_{\partial M} \sigma (x,y,z)d\Omega$,
where $\partial M$ is boundary of the molecule,  $\sigma$ the energy density and 
$d\Omega=\sqrt{g}dxdydz$. 
The energy minimization via the first variation leads to  the Euler Lagrange equation, 
\begin{equation}\label{EL}
\frac{\partial e}{\partial S} -
\frac{\partial}{\partial x} \frac{\partial e}{\partial S_x} -
\frac{\partial}{\partial y}\frac{\partial e}{\partial S_y} -
\frac{\partial}{\partial z}\frac{\partial e}{\partial S_z} 
=0,
\end{equation}
where $e=\sigma\sqrt{g}$.
For a homogeneous surface, $\sigma=\sigma_0$, a constant, Eq. (\ref{EL}) leads to
the vanishing of the mean curvature 
$\sigma_0\nabla \cdot \left(\frac{\nabla S} {\sqrt{g}}\right)=3\sigma_0 H=0$.

For a given set of atomic coordinates, we prescribe a step function initial value for $S(x,y,z)$, i.e., a 
non-zero constant $S_0$ inside a sphere of radius $\tilde{r}$ about each atom  and zero elsewhere. 
Alternatively, a Gaussian initial value can be placed around each atomic center. The value of 
$S(x,y,z)$ is updated in the iteration except for at obstacles, i.e., a set of boundary points 
given by the collection of all of the van der Waals sphere surfaces or any other desired atomic 
sphere surfaces. Here ${H}$ and $g$ can be approximated by any standard numerical methods. For 
simplicity, we use the standard second order central finite difference. Due to the stability concern, 
we choose $\varepsilon<\frac{h^2}{2}$, where $h$ is the smallest grid spacing. 
The MMS is differentiable, probe independent, and consistent with the surface free energy minimization.

\begin{figure*}[!tb] 
\begin{center}  
\begin{tabular}{ccc}   
\includegraphics[width=0.3\textwidth]{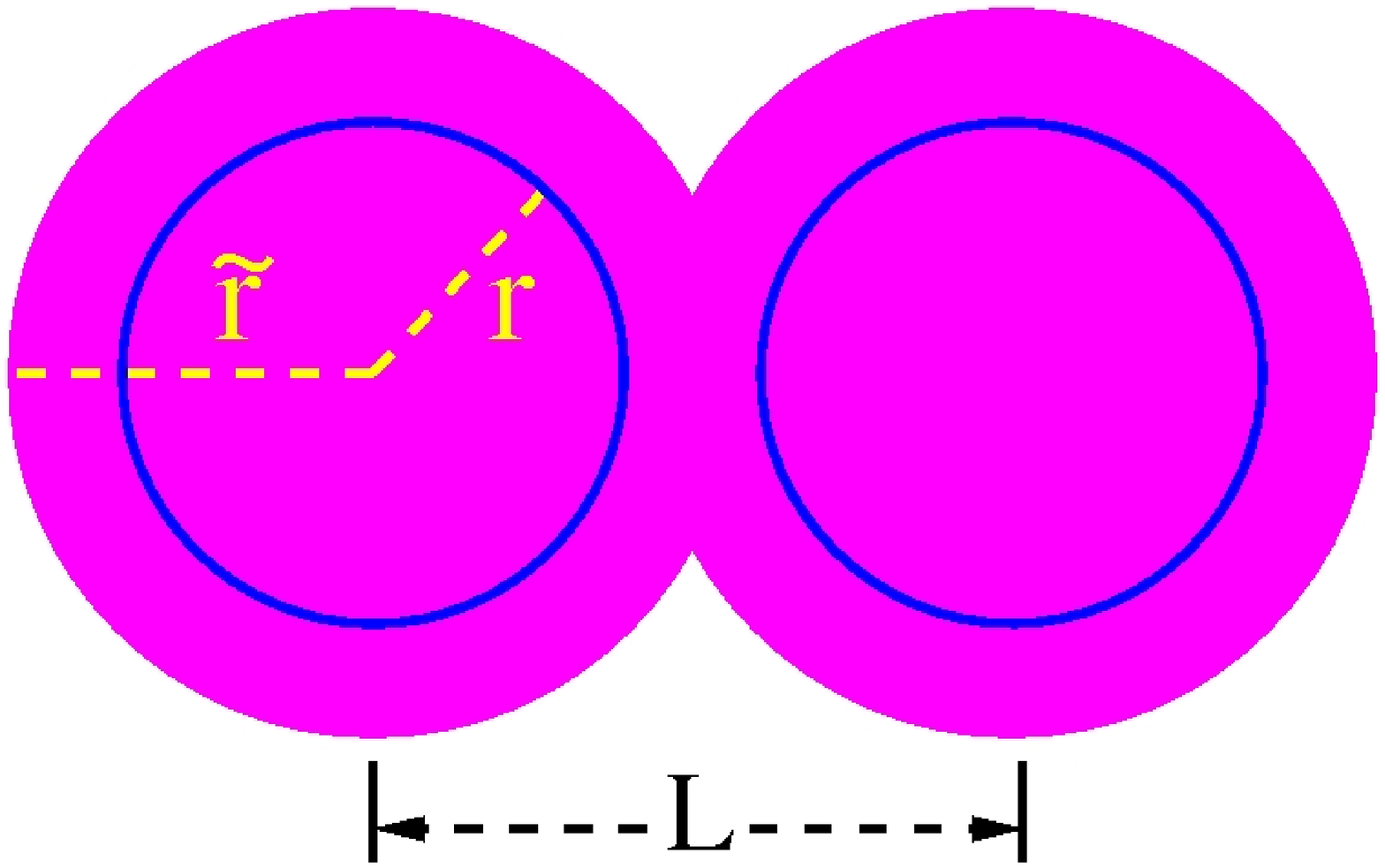}&
\includegraphics[width=0.3\textwidth]{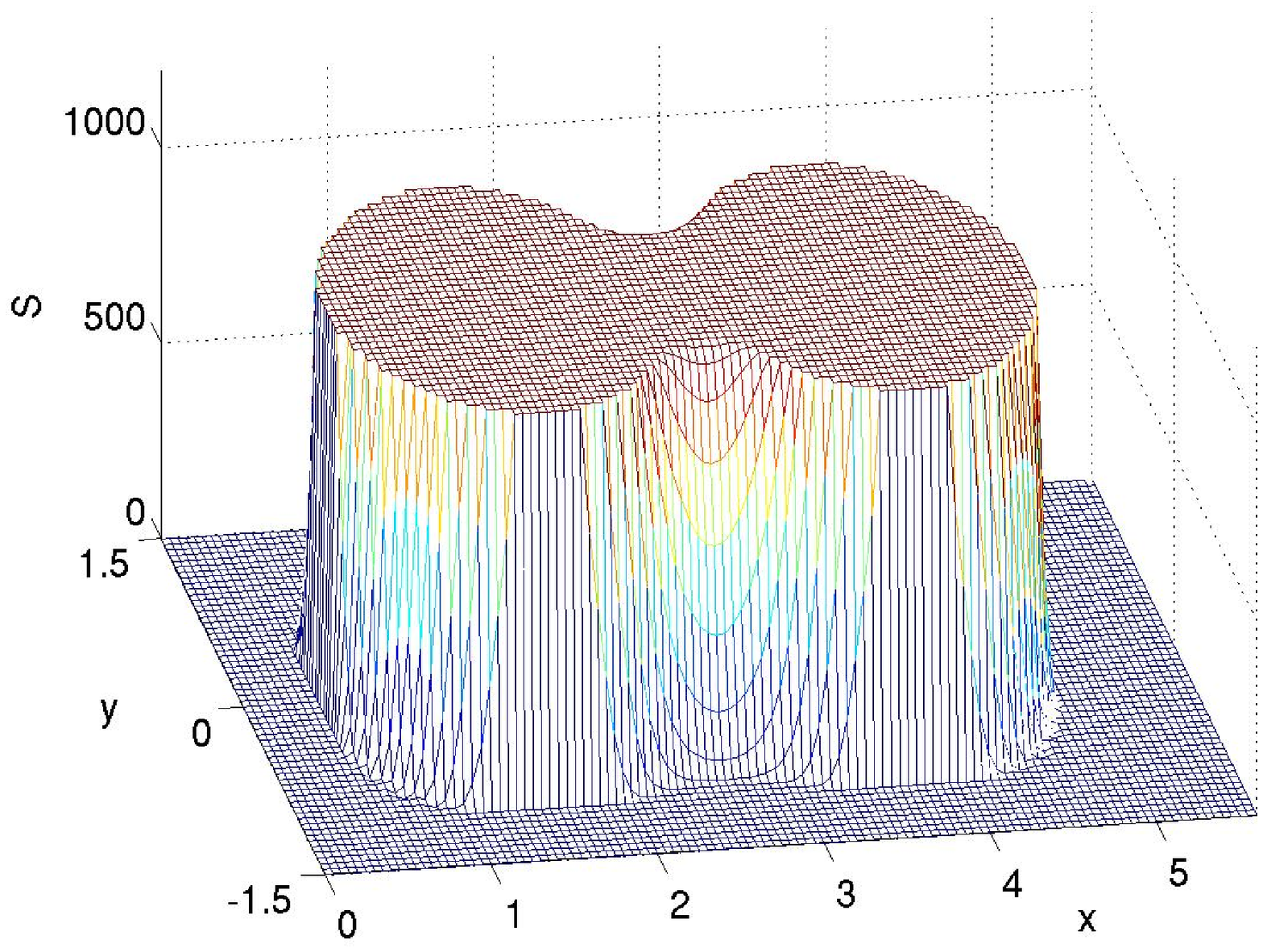}&
\includegraphics[width=0.3\textwidth]{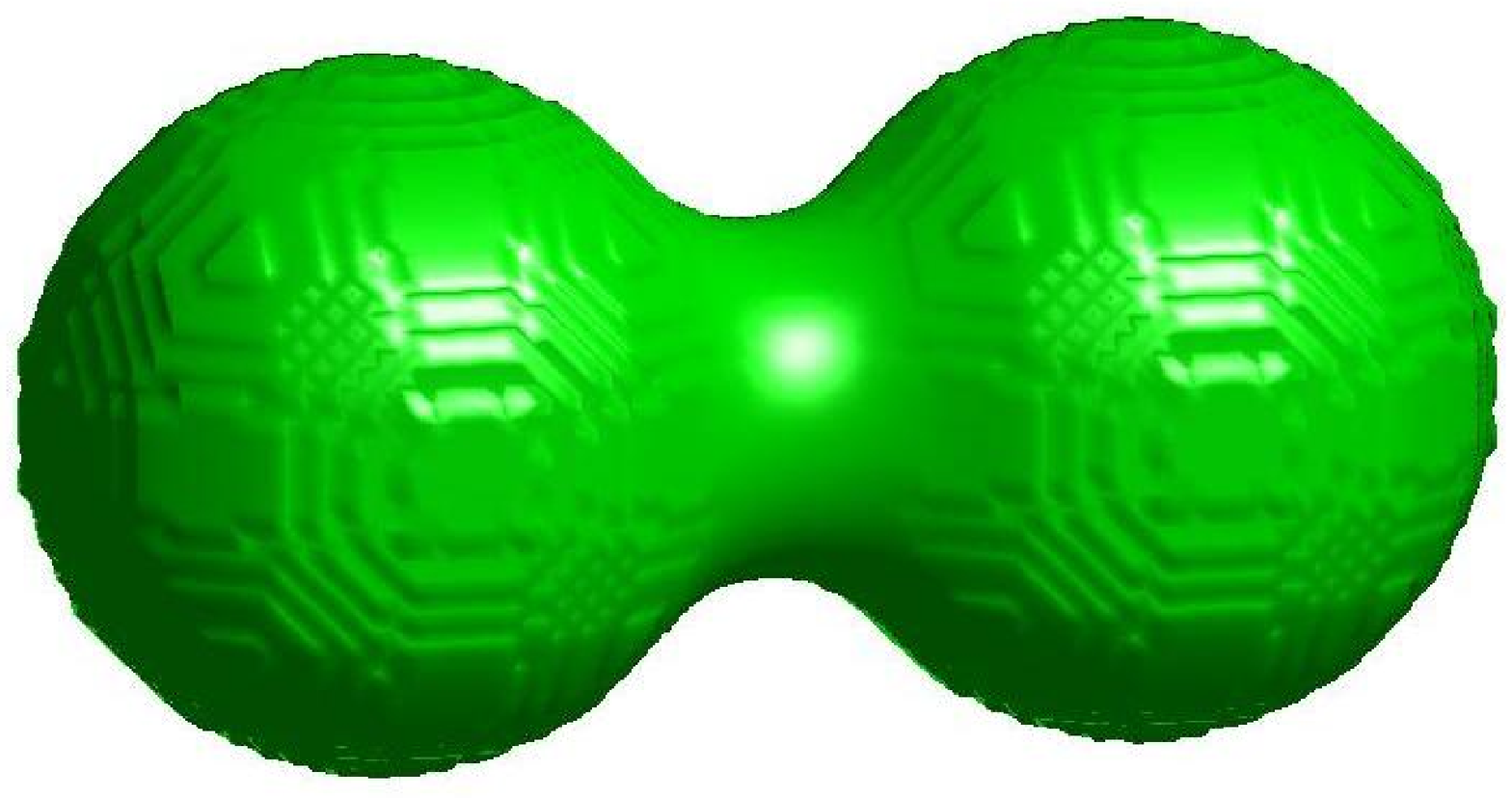}\\
(a) & (b) & (c) \end{tabular}   
\end{center}   
\caption{MMS generation. 
         (a) Illustration of $r,~ \tilde{r}$, and $L$ at a cross section $z=0$;   
         (b) $S(x,y,z=0)$ shows a family of level surfaces. 
         (c) The isosurface extracted from $S=0.99S_0$
}
\label{fig.diatom1}
\end{figure*}

As a proof of principle, we illustrate our ideas by a few examples. We first test the proposed 
method for the MMS of a diatomic molecule. The atomic radius is $r$ and their central distance 
is $L$. First, we consider a step function initial value  with $\tilde{r} > L/2 >r$, see Fig. 
\ref{fig.diatom1} (a) for an illustration. Of course, the steady state solution of $S(x,y,z)$ does not 
directly provide a surface. Instead, it gives rise to a family of level surfaces, which includes 
the desired MMS. It turns out that  $S(x,y,z)$ is very flat away from the MMS, while it
sharply varies at the MMS. In other word,  $S(x,y,z)$ is virtually a step function at the desirable 
MMS, see Fig. \ref{fig.diatom1} (b). Therefore, it is easy to extract the MMS as an isosurface at $S(x,y,z)=C$ 
as shown in Fig. \ref{fig.diatom1} (c). It is convenient to choose $C=(1-\delta)S_0$, where $\delta>0$ is a very 
small number and can be calibrated by standard tests. Computationally, by taking $S_0=1000$, satisfactory 
results can be attained by using $\delta$ values ranging from 0.004 to 0.01. We next test if there 
is any initial value constraint in our method. Indeed, it is found if $r < \tilde{r} < L/2$, two isolated 
spheres are obtained instead of the MMS. Therefore initial connectivity ($\tilde{r} > L/2$) is crucial 
for the formation of MMSs.

\begin{figure*}[!tb] 
\begin{center}  
\begin{tabular}{ccc}   
\includegraphics[width=0.3\textwidth]{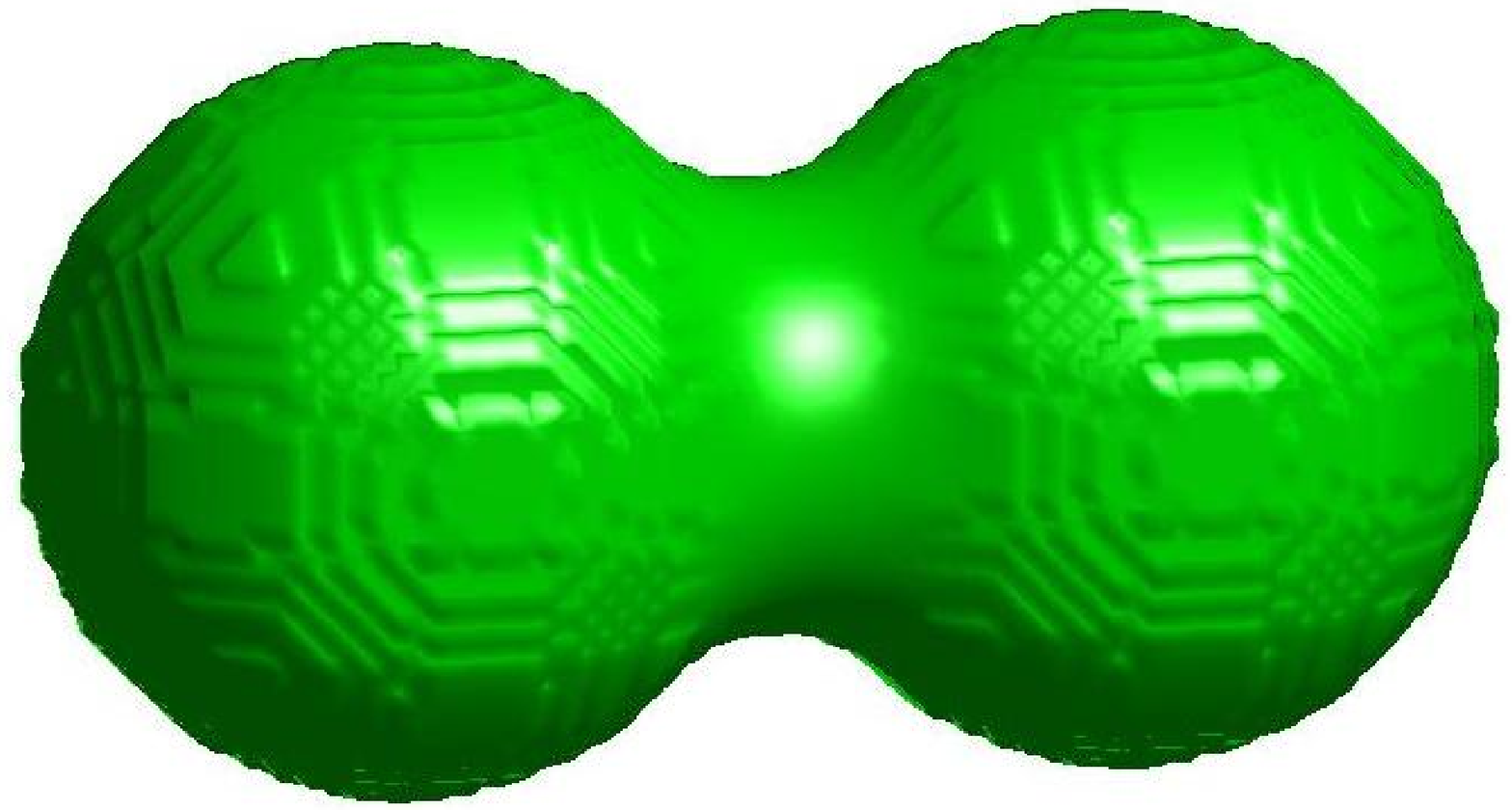}&
\includegraphics[width=0.3\textwidth]{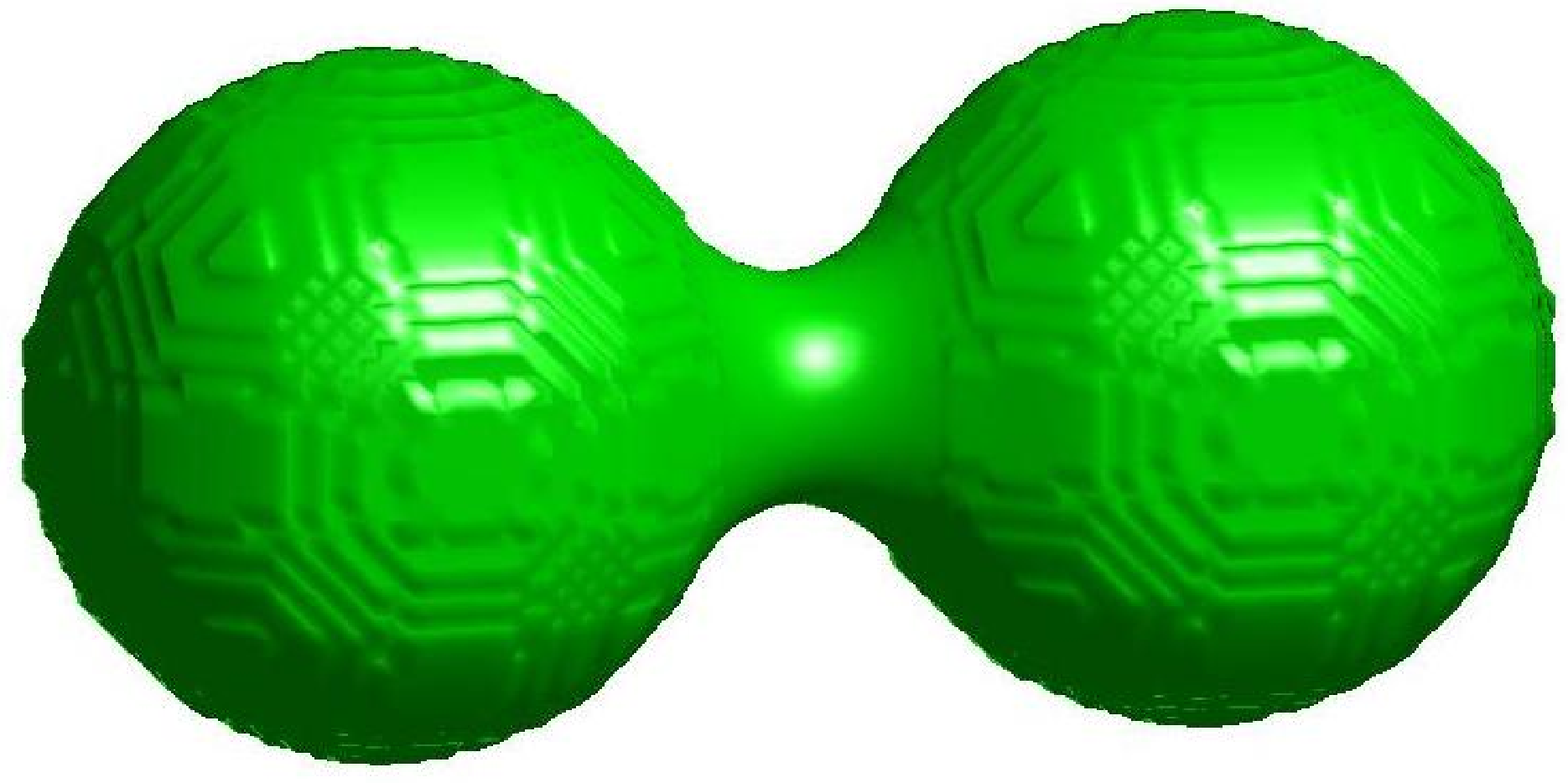}&
\includegraphics[width=0.3\textwidth]{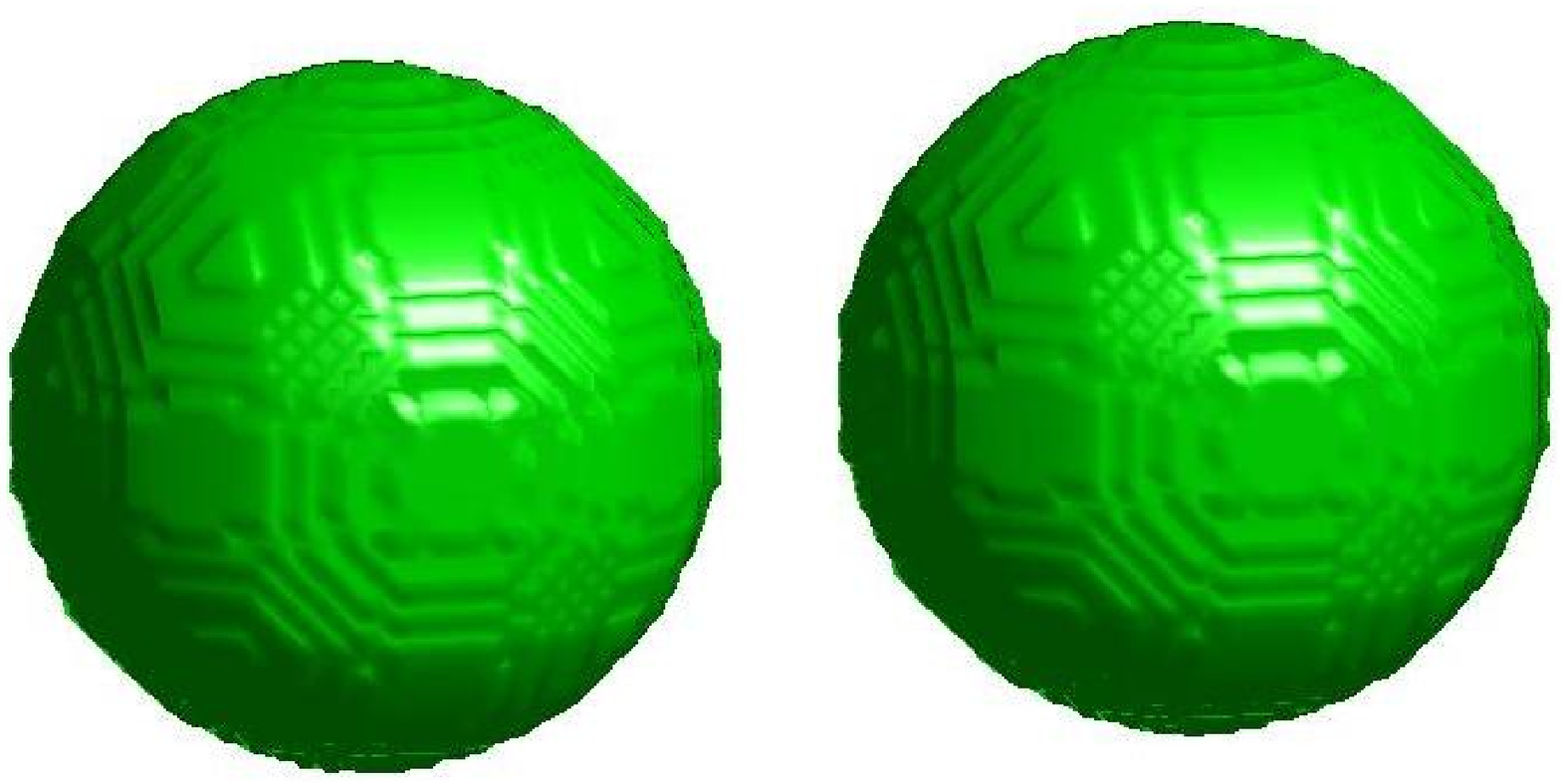}\\
(a) & (b) & (c) 
\end{tabular}   
\end{center}   
\caption{The MMS of a diatomic molecule at different separation lengths $L$.
         (a) $L=2.0r$;
         (b) $L=2.4r$;
         (c) $L=2.45r$.
}
\label{fig.diatom2}
\end{figure*}

It is important to know whether the initially connected $S(x,y,z)$ could eventually separate 
into two regions when $L$ is sufficiently large. The lower bound and the upper bound of the 
MMS areas are $4\pi r^2$ and $8\pi r^2$, respectively for the diatomic system. When $L$ is small, 
the MMS consists of a catenoid and parts of two spheres  and the MMS area is smaller than the 
upper bound, see Fig. \ref{fig.diatom2} (a). When the separation length $L$ is  gradually increased, 
the MMS area grows, while the neck of the MMS surface becomes thinner and thinner, see 
Fig. \ref{fig.diatom2} (b). At a critical distance $L_c > 2r$, the MMS area reaches the upper bound 
$8\pi r^2$ and the MMS breaks into  two disjoint pieces.
%Consistent results are obtained in numerical estimation of $L_c$ by considering different
%initial values, different $r$, and different mesh size. 
The present study predicts $L_c \simeq 2.426 r$. In fact, this result is initial-value 
independent as long as $\tilde{r} > \frac{L_c}{2}=1.213r$. We found that a Gaussian initial value gives 
the same prediction.  As the molecular surface area is proportional to the surface Gibbs free 
energy, the critical value ($L_c$) might provide an indication of the molecular disassociation 
and could be used in molecular modeling.

\begin{figure*}[!tb] 
\begin{center}  
\begin{tabular}{cccc}   
\includegraphics[width=0.2\textwidth]{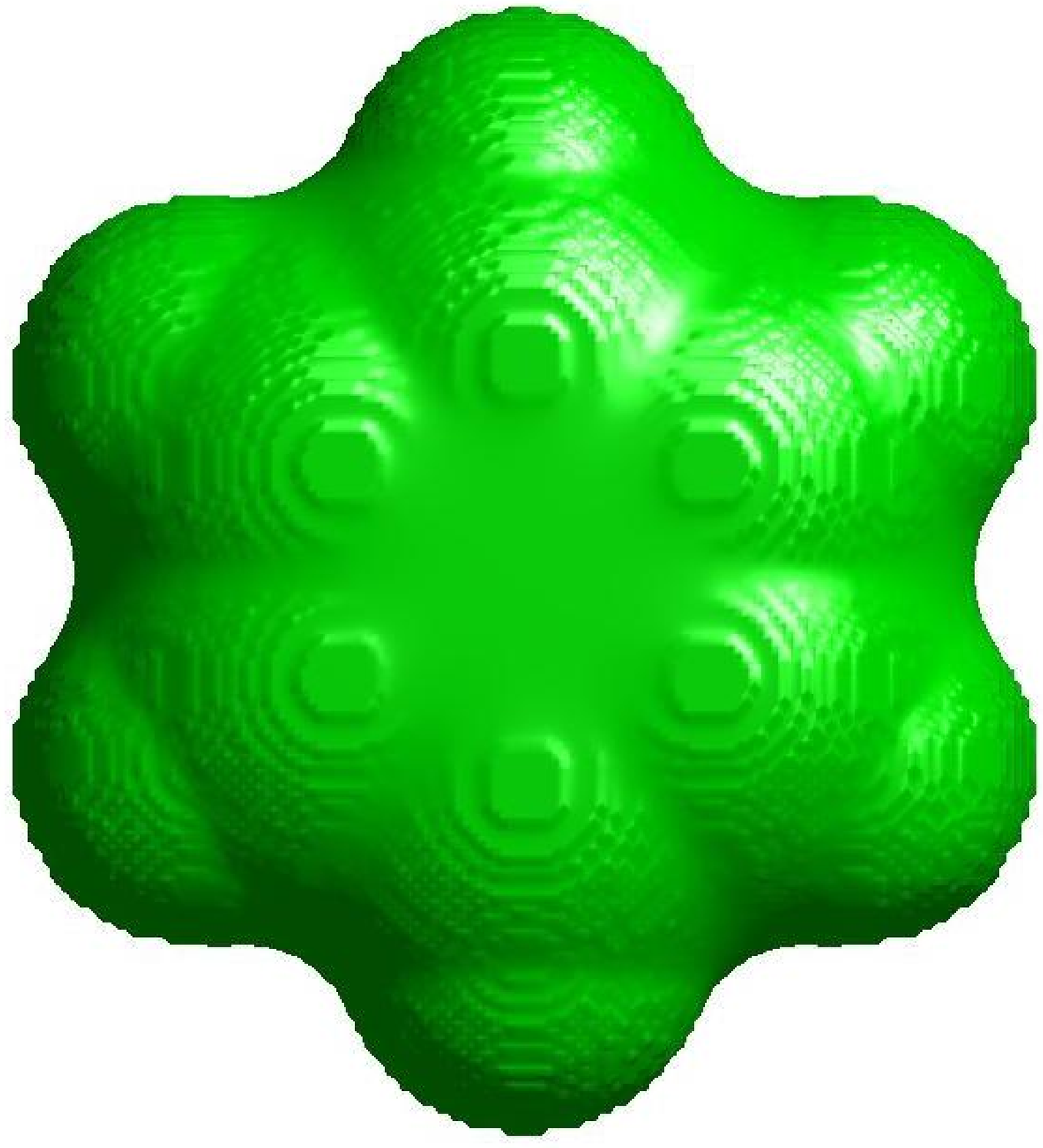} &
\includegraphics[width=0.2\textwidth]{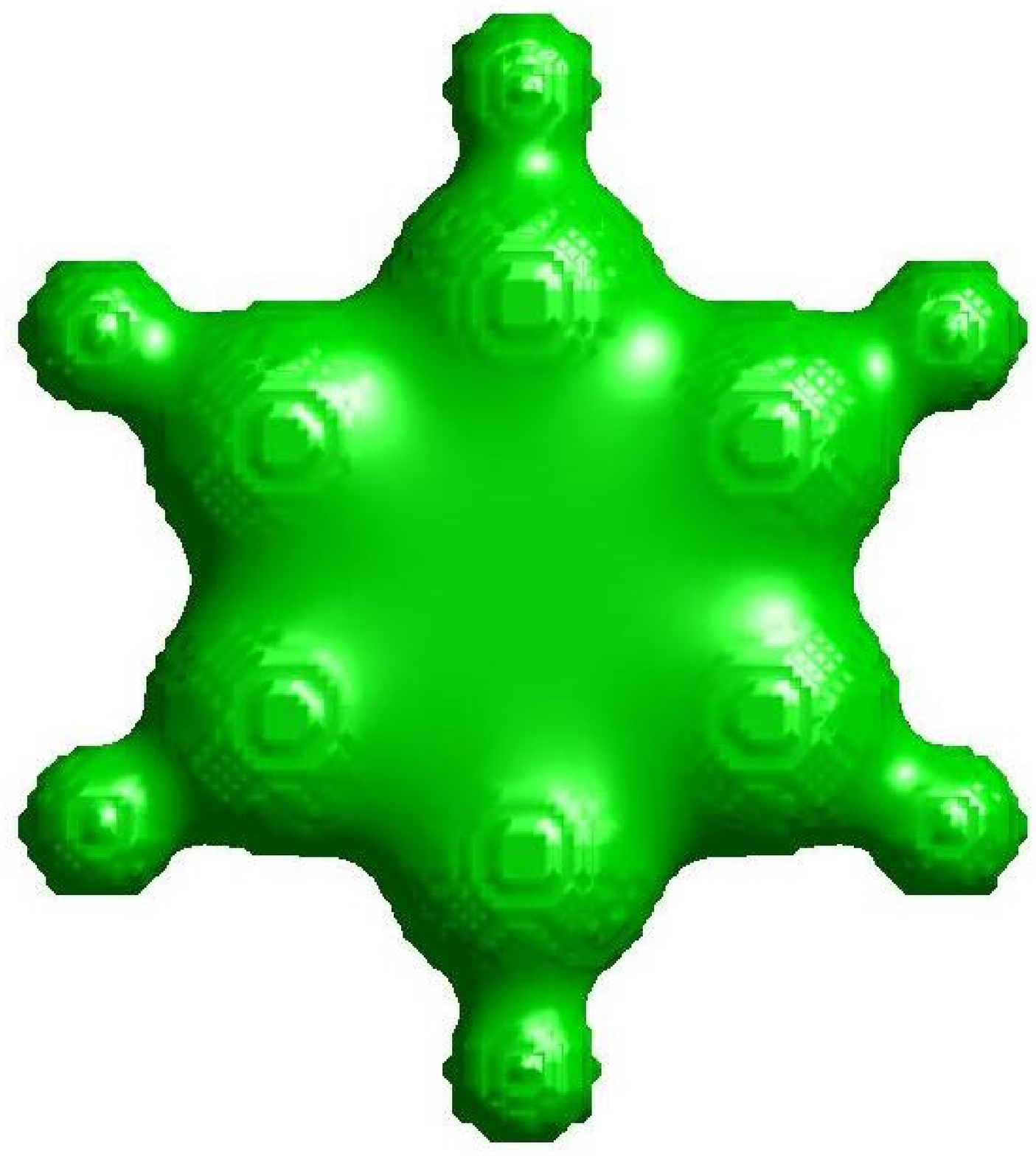} &
\includegraphics[width=0.2\textwidth]{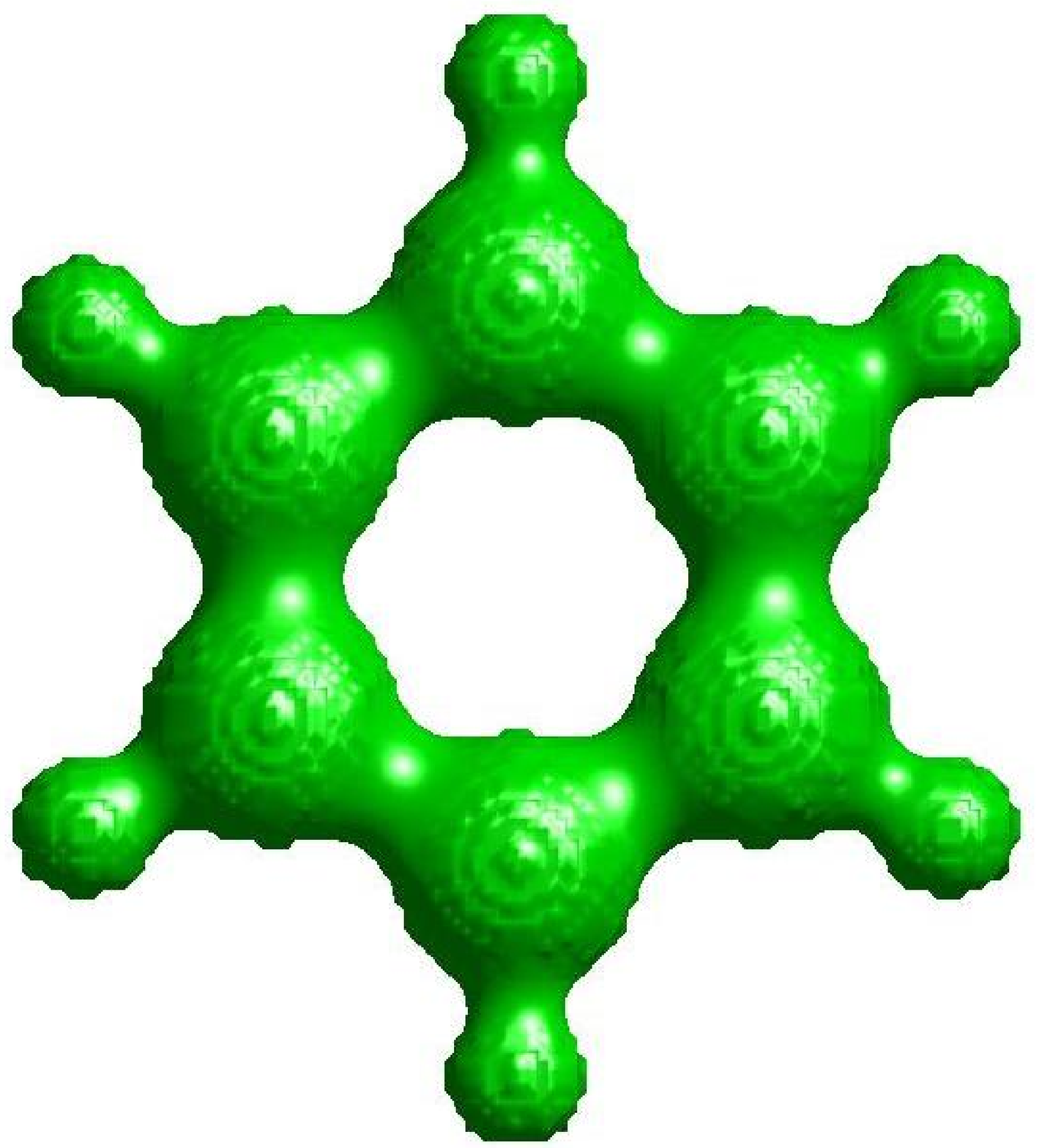} &
\includegraphics[width=0.2\textwidth]{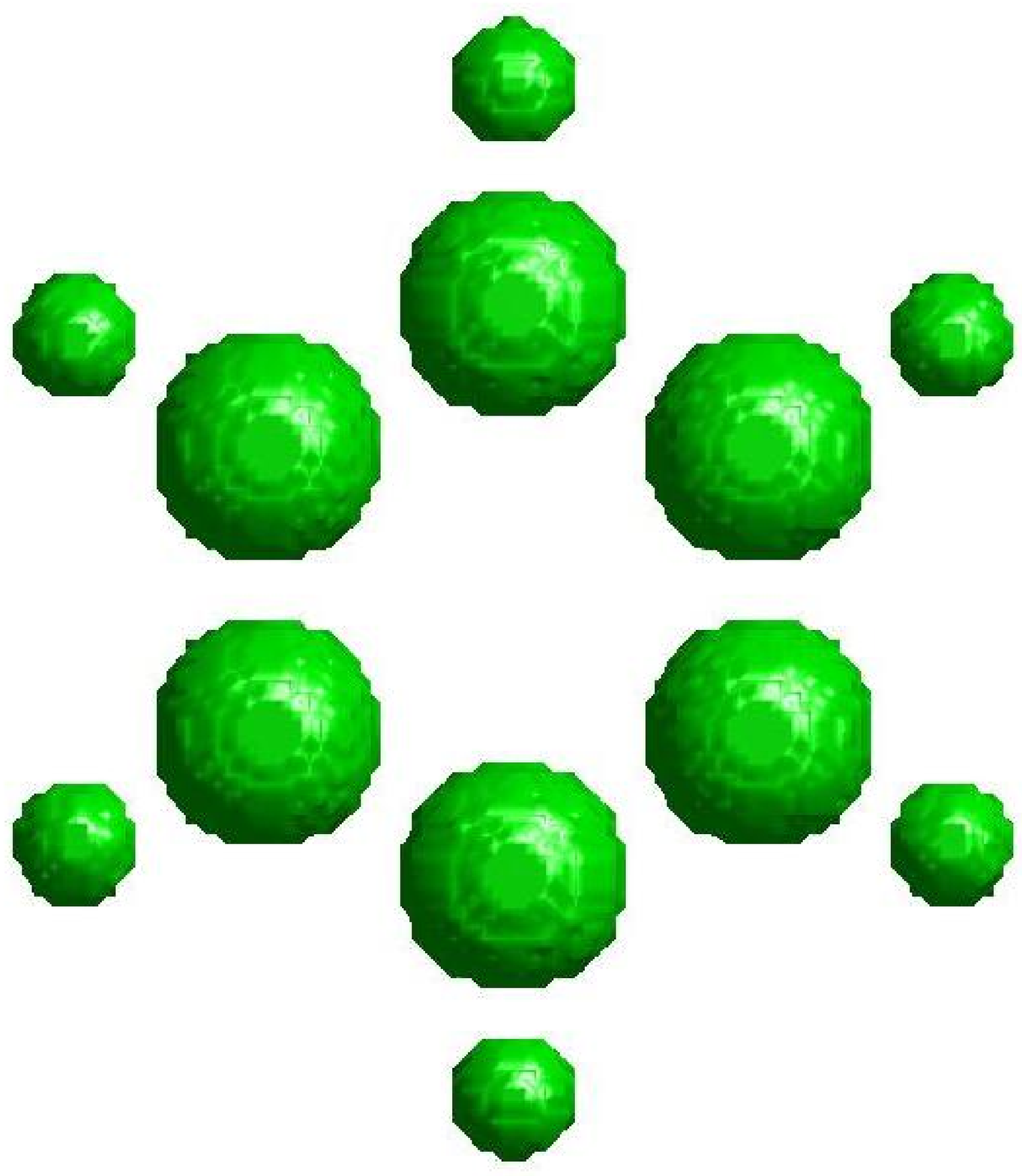} \\
(a) & (b) & (c) & (d)
\end{tabular}   
\end{center}   
\caption{The MMS of  benzene with van der Waals radii and scaled atomic radii.
(a) Van der Waals radii, $r_{\rm C}=1.7$  \AA~and $r_{\rm H}=1.2$ \AA;
(b) Atomic radii,        $r_{\rm C}=0.7$  \AA~and $r_{\rm H}=0.38$ \AA; 
(c) Scaled atomic radii, $r_{\rm C}=0.63$ \AA~and $r_{\rm H}=0.34$ \AA; 
(d) Scaled atomic radii, $r_{\rm C}=0.56$ \AA~and $r_{\rm H}=0.30$ \AA. 
}
\label{fig.ben}
\end{figure*}

We next consider the MMS of the benzene molecule which consists of six carbon atoms and six hydrogen 
atoms. The carbon atoms are in sp$^2$ hybrid states with delocalized $\pi$ stabilization. The MMSs of 
the benzene molecule with van der Waals radii ($r_{\rm vdW}$) and other atomic radii are depicted in Fig. \ref{fig.ben}.
By using the van der Waals radii, a bulky MMS is obtained. A topologically similar while smaller 
MMS is formed using the set of standard atomic radii. No ring structure is seen until the atomic radii are 
reduced by a factor of $0.9$, see Fig. \ref{fig.ben} (c).  Clearly,  all atoms are connected via catenoids. 
Eventually, the MMS decomposes into 12 pieces when radii are further reduced to slightly below their 
critical values, see Fig. \ref{fig.ben} (d). This again conforms our prediction of separation critic
$L_c \simeq 2.426 r$.

%\begin{figure*}[!tb] 
%\begin{center}  
%\includegraphics[width=0.5\textwidth]{9ant.eps} 
%\end{center}   
%\caption{The antennapedia homeodomain-DNA complex
%}
%\label{fig.9ant1}              
%\end{figure*}  

\begin{figure*}[!tb] 
\begin{center}  
\begin{tabular}{cc}   
\includegraphics[width=0.30\textwidth]{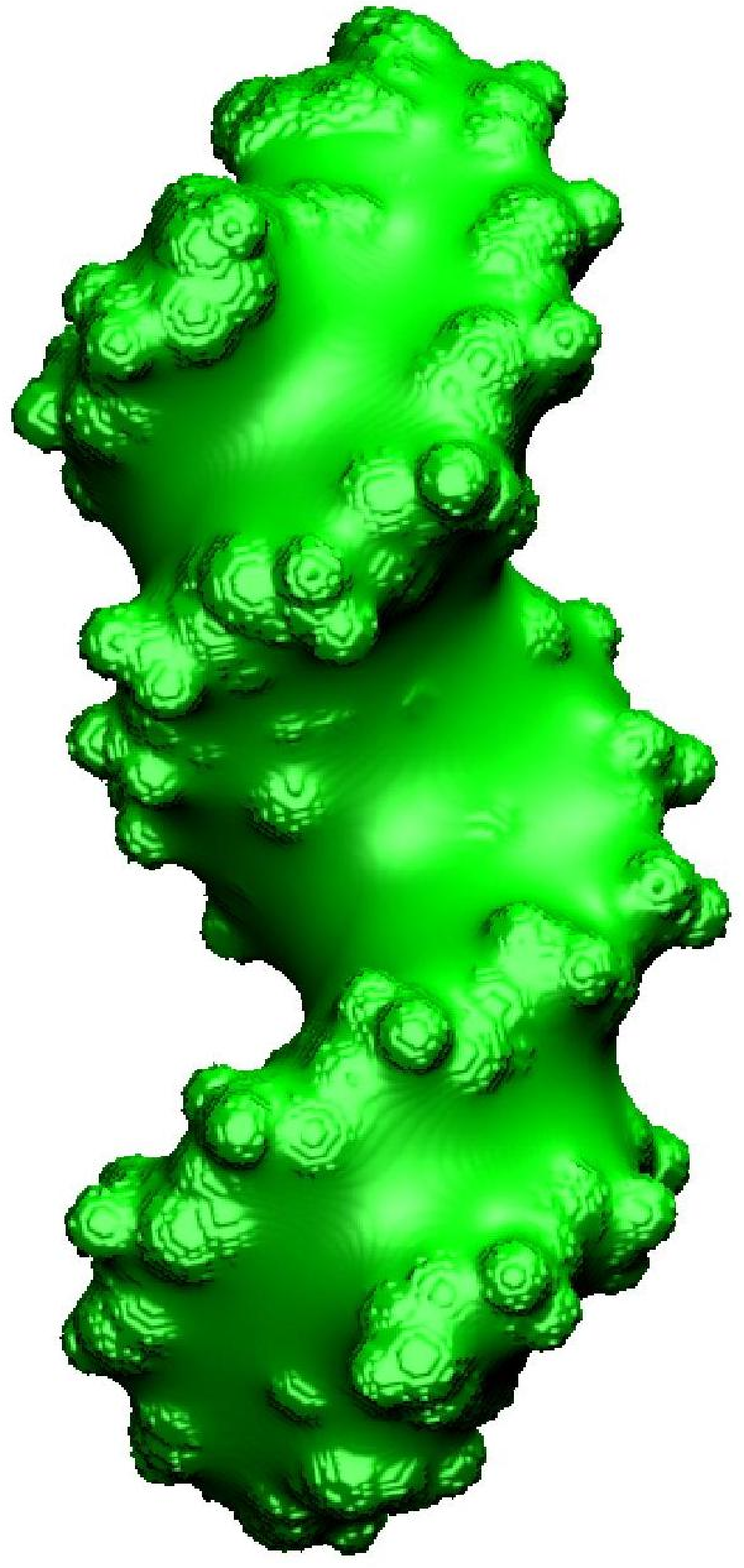}&
\includegraphics[width=0.305\textwidth]{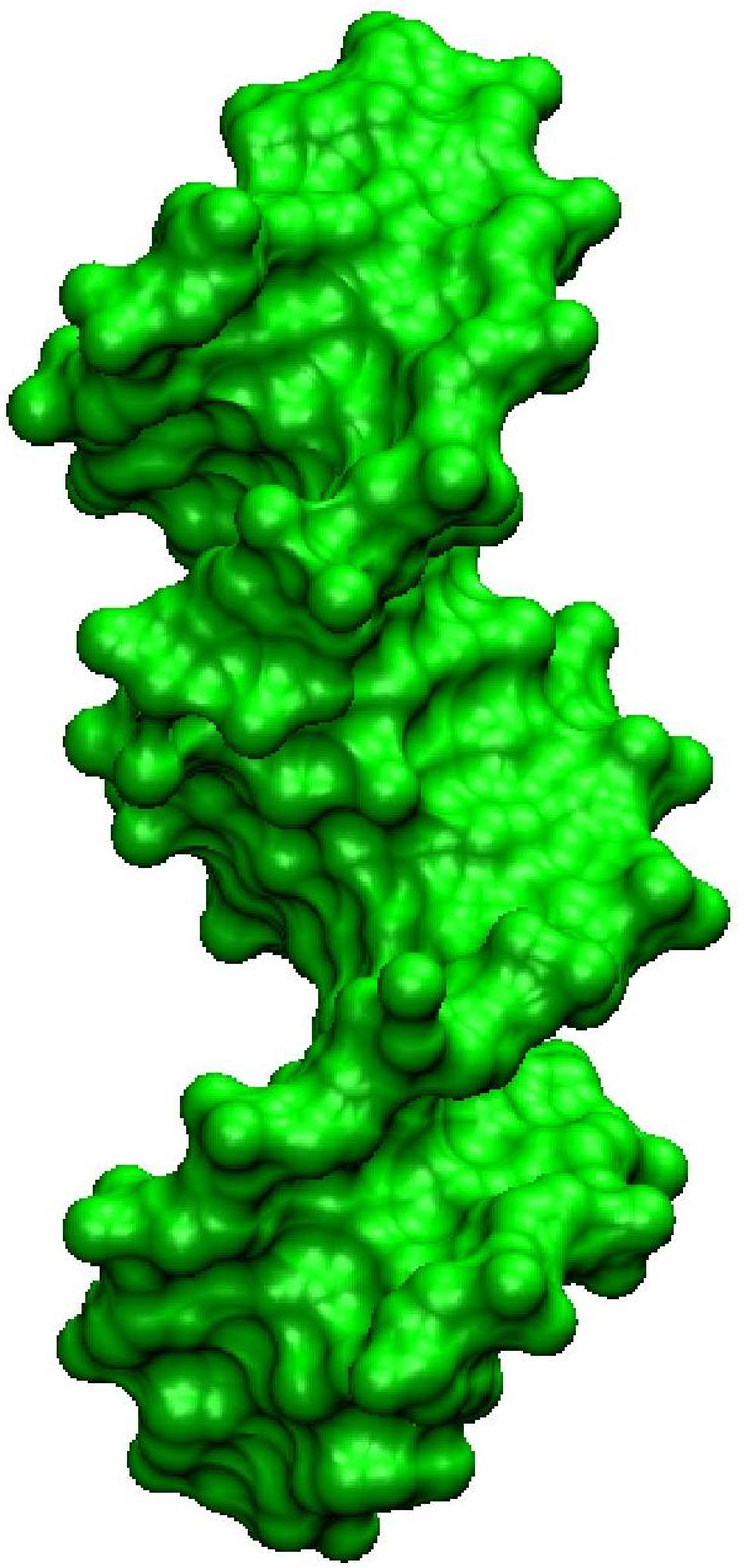}\\
(a) & (b) \\
\includegraphics[width=0.30\textwidth]{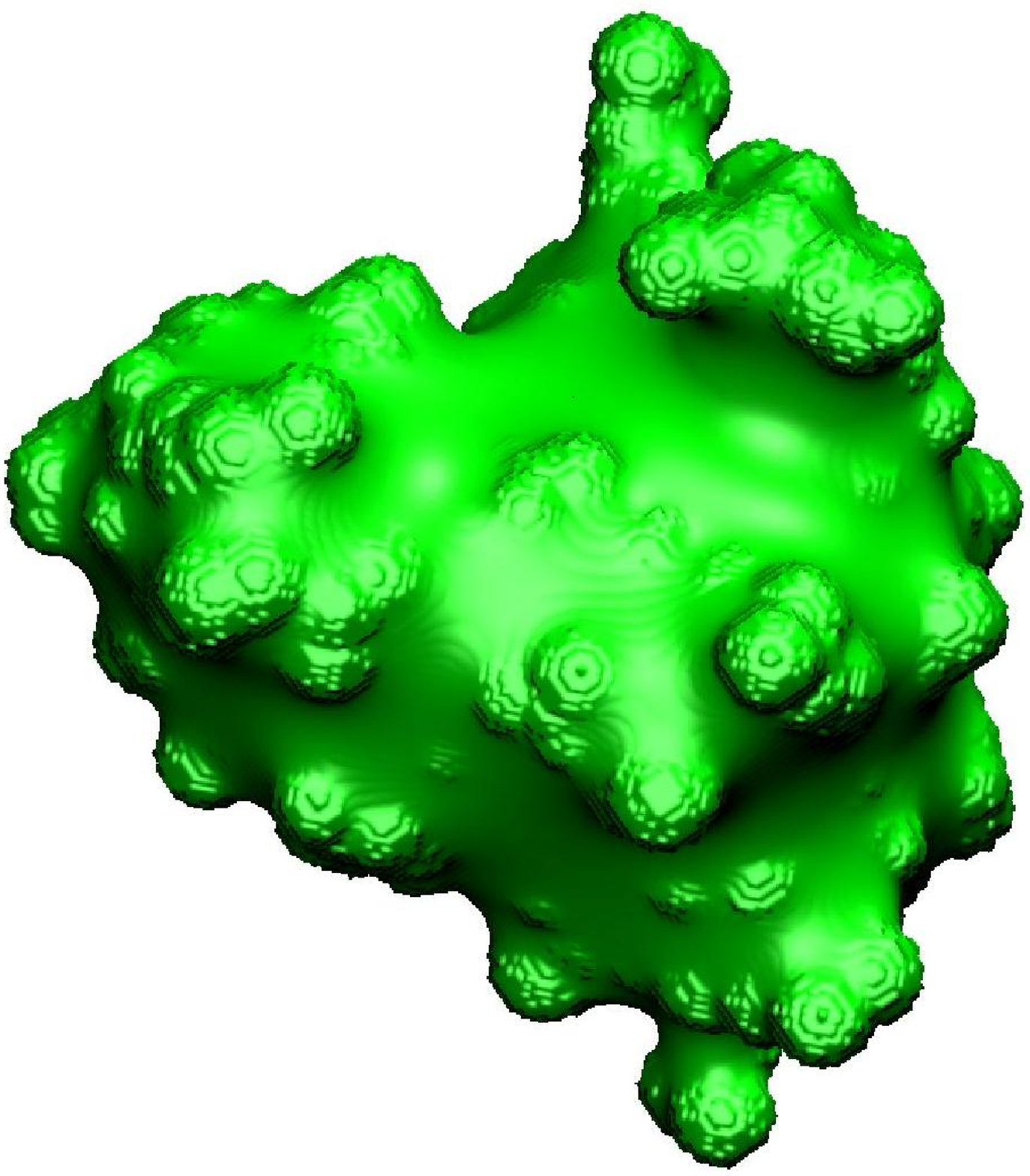}&
\includegraphics[width=0.30\textwidth]{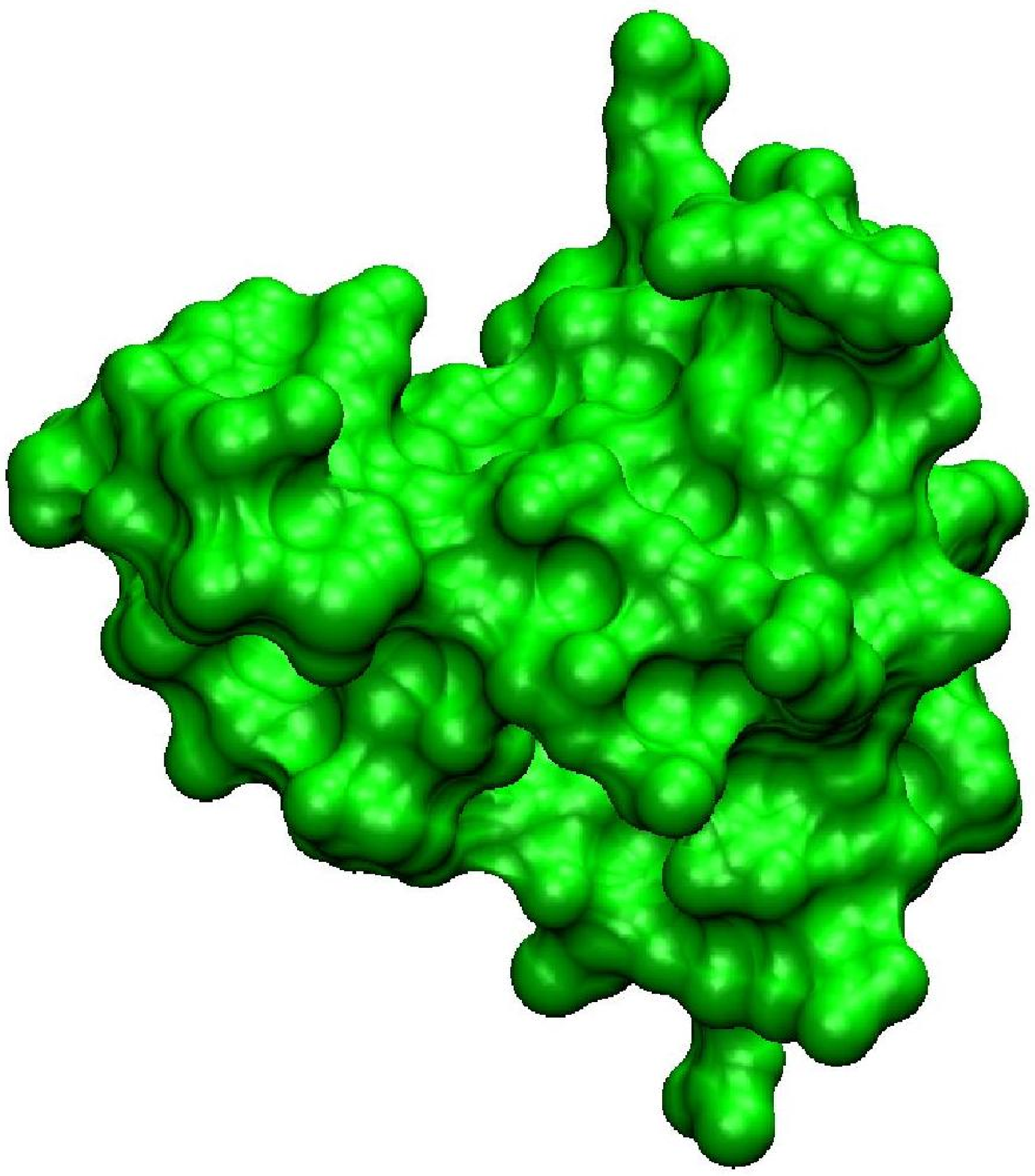}\\
(c) & (d)
\end{tabular}   
\end{center}   
\caption{
The MMS and the MS at the contacting regions of antennapedia homeodomain DNA complex
(PDB ID: 9ant). The MMSs of the DNA and the antennapedia homeodomain  
are shown in (a) and (c), respectively,
while the MS ones are shown in (b) and (d), respectively.
}
\label{fig.9ant2}              
\end{figure*}

Finally, we employ our MMS to study the mechanism of molecular recognition in 
protein-DNA interactions. NMR and molecular dynamics studies suggest that antennapedia achieves specificity 
through an ensemble of rapidly fluctuating DNA contacts \cite{Billeter}. While X-ray structure indicates a 
well-defined set of contacts due to side chains constraints \cite{Fraenkel}. % see Fig. \ref{fig.9ant1}. 
In the present work, we reveal flat contacting interfaces which stabilizing the protein-DNA 
complex. Figs. \ref{fig.9ant2}(a) and \ref{fig.9ant2}(c) depict the MMSs of 
antennapedia and DNA (PDB ID: 9ant), generated  by using $\tilde{r}=1.3r_{\rm vdW}$ and 
$h=0.2$ \AA. Clearly, the binding site of the DNA (middle groove) has a large facet,
which is absent from the top and bottom grooves of the DNA.  Interestingly, the MMS of
the protein exhibits a complimentary facet.  For a comparison, the molecular surfaces (MSs) 
generated by using the program MSMS \cite{Sanner} with the same set of van der Waals radii  and a 
probe radius of 1.5 \AA~are depicted in Figs. \ref{fig.9ant2} (b) and  \ref{fig.9ant2} (d).
Apparently, it is very difficult to recognize the complementary binding interfaces from MSs. 
It is interesting to note that the MMS also better reveals the skeleton of the DNA's 
double helix structure.  To quantitate the affinity at the contacting site, we compute
the mean distance between the MMSs of the protein and DNA by using about 7200 surface
vertices over the binding domain. A small mean distance of 0.4054 \AA~ unveils a
close contact between two facets. Relatively small standard deviation of 0.3401 \AA~ indicates
the smoothness of the contacting facets. In contrast, inconclusive mean (0.8697 \AA)~and 
standard deviation (0.5818 \AA)~ were found from the corresponding MSs.  This study indicates the 
great potential of the proposed MMS for  biomolecular binding sites prediction and recognition. 

We have introduced a novel concept, the minimal molecular surface (MMS), for 
the modeling of biomolecules, based on the speculation of free energy minimization for stabilizing 
a less polar molecule in a polar solvent. The MMS is probe independent, differentiable, and 
consistent with surface free energy minimization. A novel hypersurface approach based on the 
theory of differential geometry is developed to generate the MMSs of arbitrarily complex 
molecules. Numerical experiments are carried out on few-atom and many-atom systems to demonstrate 
the proposed method. It is believed that the proposed MMS provides a new paradigm for the studies of 
surface biology, chemistry and physics, in particular, for the analysis of stability, solubility, 
solvation energy, and interaction of  macromolecules, such as proteins, membranes, DNAs and RNAs. 
%We will examine the MMS effect in protein structure prediction and rational drug design. 
It has potential applications not only in science, but also in technology, such as
vehicle design and packaging problems.
%\newpage

\appendix
\section{Supplement: Derivation of the mean curvature evolution equation}

Consider a $C^2$ immersion $f: U\rightarrow {\Bbb R}^4 $, where $U\subset {\Bbb R}^3$ is an open set. 
Here $f(u) =(f_1(u),f_2(u),f_3(u),f_4(u))$ is a hypersurface element (or a position vector), and
$u=(u_1,u_2,u_3)\in U$. 

Tangent vectors (or directional vectors) of $f$ are $X_i=\frac{\partial f} {\partial u_i}$. The Jacobi 
matrix of the mapping $f$ is given by $Df=(X_1,X_2,X_3)$.

The first fundamental form is a symmetric, positive definite metric tensor of $f$, given by
$I:=(g_{ij})=(Df)^T\cdot(Df)$. Its matrix elements can also be expressed as $g_{ij}=<X_i,X_j>$, where 
$<,>$ is the Euclidean inner product in ${\Bbb R}^4$, $i,j =1,2,3$.

Let  $\nu(u)$ be the unit normal vector given by the Gauss map  $\nu: U\rightarrow S^3$,
\begin{equation}\label{normal}
\nu(u_1,u_2,u_3):=X_1 \times X_2 \times X_3/ \|X_1 \times X_2 \times X_3\| \in \bot_uf,
\end{equation}
where the cross product in ${\Bbb R}^4$ is a generalization of that in ${\Bbb R}^3$. 
Here $\bot_uf$ is the normal space of $f$ at point $p=f(u)$. The vector
$\nu$ is perpendicular to the tangent hyperplane $T_uf$ at $p$. Note that  
$T_uf\oplus \bot_uf =T_{f(u)}{\Bbb R}^3$, the tangent space at $p$.
By means of the normal vector $\nu$ and tangent vector $X_i$, the second  fundamental form is given by 
\begin{equation}
II(X_i,X_j)=(h_{ij})=(<-{\partial\nu\over\partial u_i}, X_j>).
\end{equation}
The mean curvature can be calculated from
\begin{equation}\label{meanc}
H=\frac{1}{3}h_{ij}g^{ji},
\end{equation}
where we use the Einstein summation convention, and $g^{ij}=g_{ij}^{-1}$.

%We define the shape operator of $f$ as the Weingarten map:  $L:=-D\nu \circ (Df)^{-1}$.  
%Since $L$ is a self-adjoint operator, we have
%\begin{equation}
%I(LX_i,X_j)=(<-{\partial\nu\over\partial u_i}, X_j>)
%=(<\nu, {\partial^2 f\over \partial u_i\partial u_j}>)
%=II(X_i,X_j)=(h_{ij}),
%\end{equation}
%The third and fourth  fundamental forms are conveniently given in terms of the shape operator
%\begin{eqnarray}
%III(X_i, X_j)&=&I(L^2X_i,X_j)=(e_{ij})=(<{\partial \nu\over \partial u_i},{\partial \nu\over \partial u_j}>)\\
%IV(X_i, X_j)&=&I(L^3X_i,X_j).
%\end{eqnarray}
 
%Principal curvatures are defined as the eigenvalues of $L$ with eigenvectors being unit tangent vectors.
%Appropriate organization of the principal curvatures gives rise to the first three mean curvatures
%\begin{eqnarray}
%K_1&=&\frac{1}{3}(\kappa_1+\kappa_2+\kappa_3)\\
%K_2&=&\frac{1}{3}(\kappa_1\kappa_2+\kappa_1\kappa_3+\kappa_2\kappa_3)\\
%K_3&=&\kappa_1\kappa_2\kappa_3
%\end{eqnarray}
%where $K_1=H=\frac{1}{3}{\rm Tr}(L)$ is the standard mean curvature and $K_3=K={\rm Det}(L)$ is the Gauss 
%curvature. The local property of the Gauss curvature is used to classify the point as elliptic, hyperbolic,
%parabolic, etc. It follows from the Cayley-Hamilton theorem \cite{Gray} that  the first four fundamental 
%forms satisfy: $IV-3HIII+3K_2II-KI=0$.

Let $U \subset {\Bbb R}^3$ be an open set and suppose $\overline{U}$ is compact with boundary $\partial U$.
Let $f_\varepsilon:\overline{U}\rightarrow {\Bbb R}^4$ be a family of hypersurfaces indexed by $\varepsilon>0$,
obtained by  deforming $f$ in the normal direction according to the mean curvature. Explicitly, we set 
\begin{equation}\label{evol}
f_\varepsilon (x,y,z):=f(x,y,z)+\varepsilon H \nu(x,y,z).
\end{equation}
We wish to iterate this leading to a minimal hypersurface, that is  $H=0$ in all of $U$, except possibly 
where barriers (atomic constraints) are encountered.

For our purpose, let us choose $f(u)=(x,y,z,S)$, where $S(x,y,z)$ is a function of interest. 
We have the first fundamental form:
\begin{equation}
(g_{ij})=\left( 
       \begin{array}{lll}
   1+S_x^2  & S_xS_y  & S_xS_z \\
   S_xS_y   & 1+S_y^2 & S_yS_z \\
   S_xS_z   & S_yS_z  & 1+S_z^2 
       \end{array}\right).
\end{equation}
The inverse matrix of $(g_{ij})$ is given by
\begin{equation}
(g^{ij})=\frac{1}{g}\left( 
       \begin{array}{lll}
   1+S_y^2+S_z^2  & -S_xS_y  &- S_xS_z \\
   -S_xS_y   & 1+S_x^2+S_z^2 &- S_yS_z \\
   -S_xS_z   & -S_yS_z  & 1+S_x^2+S_y^2 
       \end{array}\right),
\end{equation}
where $g={\rm Det}(g_{ij})=1+S_x^2+S_y^2+S_x^2$ is the Gram determinant. 
The normal vector can be computed from Eq. (\ref{normal})
\begin{equation}
\nu=(-S_x,-S_y,-S_z,1)/\sqrt{g},
\end{equation}
The second fundamental 
form is given by
\begin{equation}
(h_{ij})=\left(\frac{1}{\sqrt{g}}S_{x_ix_j}\right),
\end{equation}
i.e., the Hessian matrix of $S$.

We consider a family $f_\varepsilon=(x,y,z,S_\varepsilon)$, where
\begin{equation}\label{evolu}
S_\varepsilon(x,y,z)=S(x,y,z)+\varepsilon H \frac{1}{\sqrt{g}}.
\end{equation}
The explicit form for the mean curvature can be written as
\begin{equation}
H=\frac{1}{3} \nabla \cdot \left(\frac{\nabla S} {\sqrt{g}}\right).
\end{equation}
 Thus, we arrive at the final evolution scheme
\begin{equation}\label{evolut2}
S_\varepsilon(x,y,z)=S(x,y,z)+ \frac{\varepsilon}{3\sqrt{g}} 
\nabla \cdot \left(\frac{\nabla S} {\sqrt{g}}\right).
\end{equation}

To balance the growth rate of the mean curvature operator, we replace $H$ by $gH$, 
which is permissible since  $g$ is nonsingular. This leads to Eq. (\ref{evolut}) of the 
main text.


\begin{thebibliography}{99} 
\newcommand{\ncAddPaper} [7]{\bibitem{#1}#2,  #3, {\it #4},    {\bf #5}, #6 (#7).} 
\newcommand{\ncAddBook}  [5]{\bibitem{#1}#2,  #3, {\it #4},    (#5).} 
\newcommand{\ncAddBookC}  [6]{\bibitem{#1}#2,  #3, {\it #4},   #5 (#6).} 
\newcommand{\ncAddPaperC}[6]{\bibitem{#1}#2,  #3, {\it #4},    { #5}, (#6). } 
\newcommand{\ncAddProced}[6]{\bibitem{#1}#2,  #3, {\it in #4}, {\bf #5}, #6.} 

%\ncAddPaper{CoreyPauling}{R.B. Corey  and L. Pauling} 
%{Molecular models of amino acids, peptides and proteins} {Rev. Sci. Instr.} {24}{ 621-627}{1953}


 
\ncAddPaper{Kuhn}{L.A.  Kuhn, M. A. Siani, M. E. Pique, C. L. Fisher, E. D. Getzoff and J. A. Tainer}
     {The interdependence of protein surface topography and bound water molecules
       revealed by surface accessibility and fractal density measures} 
     {J. Mol. Biol.} {228} {13-22} {1992}


\ncAddPaper{Richards}{F.M. Richards} {Areas, volumes, packing and protein structure}
        {Annu. Rev. Biophys. Bioeng.} {6} {151-176} {1977}   

\ncAddPaper{Connolly} {M.L. Connolly} {Analytical molecular surface calculation.}
 {J. Appl. Crystallogr.} {16}  {548-558} {1983}

%\ncAddPaper{Lee} {B.  Lee and F.M.  Richards} 
%{Interpretation of protein structures: estimation of static accessibility}
%{J. Mol. Biol.}  {55} {379-400} {1973}   


\ncAddPaper{Spolar} {R.S. Spolar and M.T. Jr. Record} 
          {Coupling of local folding to site-specific binding of proteins to DNA}
               {Science} {263} {777-184} {1994}   

\ncAddPaper{Crowley}{P.B. Crowley and A. Golovin}
    {Cation-pi interactions in protein-protein interfaces} 
   {Proteins - Struct. Func. Bioinf.} {59} {231-239} {2005} 


\ncAddPaper{Bergstrom}{C.A.S. Bergstrom, M. Strafford, L. Lazorova, A. Avdeef, K. Luthman and 
   P. Artursson} 
   {Absorption classification of oral drugs based on molecular surface properties} 
    {J. Medicinal Chem.} {46} { 558-570} {2003} 

\ncAddPaper{Dragan}{A.I. Dragan, C.M. Read, E.N. Makeyeva, E.I. Milgotina, M.E.A. 
   Churchill, C. Crane-Robinson and P.L. Privalov}
  {DNA binding and bending by HMG boxes: Energetic determinants of specificity} 
   {J.  Mol. Biol.} {343} {371-393} {2004} 

\ncAddPaper{Jackson} {R.M. Jackson and M.J. Sternberg} 
 {A continuum model for protein-protein interactions: application to the docking problem }
      {J. Mol. Biol.} {250} {258-275} {1995} 

\ncAddPaper{LiCata} {V.J. LiCata and N.M.  Allewell} 
  {Functionally linked hydration changes in Escherichia coli aspartate transcarbamylase and 
     its catalytic subunit}
    {Biochemistry}  {36} {10161-10167} {1997}    

\ncAddPaper{Raschke} {T.M. Raschke,  J. Tsai and M. Levitt} {Quantification of the hydrophobic 
  interaction by simulations of the aggregation of small hydrophobic solutes in water}
        {Proc. Natl. Acad. Sci. USA} {98} {5965-5969} {2001} 


\ncAddPaper{Das} {B.  Das and H. Meirovitch} {Optimization of solvation models for predicting
   the structure of surface loops in proteins} {Proteins} {43} {303-314} {2001}     

\ncAddPaper{WarWat} {J. Warwicker and H.C. Watson}
   {Calculation of the electric-potential in the active-site
    cleft due to alpha-helix dipoles}
    {J. Mol. Biol.} {154} {671-679} {1982}


\ncAddPaper{Honig95} {B. Honig and A. Nicholls}
{Classical electrostatics in biology and chemistry}
{Science} {268} {1144-1149} {1995}


\ncAddPaper{Andersson}{S. Andersson, S.T. Hyde, K. Larsson and S. Lind}
{Minimal surfaces and structures   from inorganic and metal crystals to cell membranes and bio polymers}
{Chem. Rev.}{88}{221-242}{1998}


\ncAddPaper{Anderson}{M.W. Anderson, C.C. Egger, G.J.T. Tiddy, J.L. Casci and K.A. Brakke} 
{A new minimal surface and the structure of mesoporous silicas} 
{Angew. Chem. Int. Ed.}{44} {3243-3248}{2005} 



\ncAddPaper{Pociecha} {D. Pociecha, E. Gorecka, N. Vaupotic, M. Cepic and J.  Mieczkowski} 
{Spontaneous breaking of minimal surface condition: Labyrinths in free standing smectic films} 
{Phys. Rev. Lett.}{95} {No. 207801}{2005} 


\ncAddPaper{Seddon} {J.M. Seddon and R.H. Templer} {Cubic phases of self-assembled amphiphilic aggregates}
{Philos. T. Royal Soc. London Ser. A-Math. Phys. Engng. Sci.} {244} {377-401} {1993}



\ncAddPaper{ChenEHOY}{Chen BL, Eddaoudi M, Hyde ST, O'Keeffe M, Yaghi OM } 
{Interwoven metal-organic framework on a periodic minimal surface with extra-large pores} 
{Science} {291}  {1021-1023}{ 2001}


\ncAddPaper{Koh}{E. Koh and T. Kim}{Minimal surface as a model of beta-sheets}
 {Prot. Struct. Func. Bioinf.} {61} {559-569} {2005} 

\ncAddPaper{Falicov} {A. Falicov and F.E.  Cohen}
{A surface of minimum area metric for the structural comparison of proteins} 
{J. Mole. Biol.} { 258} { 871-892 }{1996}

\ncAddPaper{Chopp}{D.L. Chopp}
    {Computing minimal-sufaces via level set curvature flow}
      {J. Comput. Phys.}{106}{77-91}{1993}

\ncAddPaper{Cecil}{T. Cecil}
{A numerical method for computing minimal surfaces in arbitrary dimension} 
{J. Comput. Phys.} {206} {650-660} {2005} 


\ncAddBook{Gray}{A. Gray}
{Modern Differential Geometry of Curves and Surfaces with Mathematica}
 {Second Edition} {CRC Press, Boca Raton, 1998}


%\ncAddPaper{Chen}{ S. Chen, J. Jancrick, H. Yokota, R. Kim and  S.-H. Kim}  
%{Crystal structure of a protein associated with cell division from Mycoplasma
%pneumoniae (GI: 13508053): a novel fold with a conserved sequence motif}
%{Proteins} {55}  {785-791}  {2004}  


\ncAddPaper{Billeter} {M. Billeter}  {Homeodomain-type DNA recognition} 
{Progr. Biophys. Mol. Biol.} {66} {211-225} {1996} 

\ncAddPaper{Fraenkel}{E. Fraenkel  and C.O. Pabo}
{Comparison of X-ray and NMR structures for the Antennapedia homeodomain/DNA complex}
{Nature Struc. Mol. Biol.} {5} {692 - 697} {1998}


\ncAddPaper{Sanner} {M.F. Sanner, A.J. Olson and J.C. Spehner} 
   {Reduced surface: An efficient way to compute molecular surfaces} 
   {Biopolymers} {38} {305-320} {1996}
\end{thebibliography}
\end{document}